# Polymer multimode waveguide bend based on a multilayered Eaton lens

S. HADI BADRI,[1] H. RASOOLI SAGHAI, [2] HADI SOOFI[3]

[1] Department of Electrical Engineering, Azarshahr Branch, Islamic Azad University, Azarshahr, Iran
[2] Department of Electrical Engineering, Tabriz Branch, Islamic Azad University, Tabriz, Iran
[3] School of Engineering-Emerging Technologies, University of Tabriz, Tabriz 5166616471, Iran
sh.badri@iaut.ac.ir

**Reducing the bending radius of low-index contrast waveguides is essential in reducing the size of the integrated optical components. A polymeric multimode waveguide bend is presented based on the Eaton lens. The ray-tracing calculations are utilized to truncate the Eaton lens in order to improve the performance of the bend. The truncation of the lens decreases the footprint of the bend as well. The designed waveguide bend with a radius of 18.4 μm is implemented by concentric cylindrical multilayer structure. The average bend losses of 0.69 and 0.87 dB are achieved for the $TM_0$ and $TM_1$ modes in the C-band of optical communication, respectively. The bend loss is lower than 1 dB in a bandwidth of 1520-1675 nm for both modes.**

## 1. Introduction

Polymers, due to their transparency at optical communication bands, mechanical flexibility, biocompatibility, and low refractive indices in the range of silica fibers, play an increasingly important role in passive photonic components [1]. Moreover, low-index contrast waveguides such as polymeric waveguides have low fiber coupling and propagation losses. Reducing the bending radius while maintaining a low loss is critical in densely integrated planar lightwave circuits (PLCs) which have attracted a great interest in recent years due to their low cost and ease of fabrication [2]. However, low-index contrast waveguide bends occupy a large footprint due to the small index difference between the waveguide's core and cladding. The bending loss depends on the waveguide's index-contrast, mode power distribution inside the core, and waveguide's dimensions. Moreover, the bending loss increases for higher-order modes [3]. Reducing the bending radius of low-index contrast waveguides is essential in reducing the overall footprint of the components. The radiation loss in bent waveguides should also be minimized to reduce crosstalk to neighboring waveguides. Only limited research has focused on designing polymeric multimode low-index contrast waveguide bends. Air trench placed at the corner of the single-mode waveguide bend has been proposed as bending mechanism which relies on total internal reflection. The measured bend loss for a 45° bend is 0.16 dB at 1550 nm. However, the 63×63 μm² footprint is considerably large for a 45° bent waveguide [4]. Multimode tapered polymer waveguide bends with bending radii in the range of millimeter have been proposed [5]. Other polymeric single-mode or multimode waveguide bends have also been reported with bending radii in the millimeter range [3, 6, 7]. Different bending mechanisms have been studied for silica waveguides such as multilayer air trenches [8] and gradient refractive index (GRIN) structures [9, 10]. For single-mode and multimode high-index contrast waveguides, various methods such as mode converters [11, 12], transformation optics [13, 14], and inverse design [15] have been utilized.

In this paper, we theoretically design a polymeric multimode waveguide bend based on the truncated Eaton lens and numerically evaluate its performance. Recently, interesting applications have been introduced for GRIN lenses such as the Maxwell's fisheye [16, 17], Luneburg [18, 19], and Eaton [10] Lenses. In our previous work, we implemented the Eaton lens by graded photonic crystal (GPC) for bending a single-mode silica waveguide [10] while in this work we implement the Eaton lens based on multilayer structure for bending a polymeric multimode waveguide. The numerical evaluations prove that the designed bent waveguide has a promising performance with a low footprint compared to previously designed bends.

## 2. Bend design

In this section, we design a bend for a polymeric waveguide with a width of $w_{WG} = 5.5\ \mu m$. The refractive indices of the core and cladding materials are chosen to be $n_{core}$=1.4816 and $n_{clad}$=1.4625, respectively [4]. The Eaton lens can bend light in 90, 180, and 360 degrees. The refractive index distribution of the Eaton lens bending light by 90° is described by [20]:

$$n_{lens}^2 = \frac{R_{lens}}{n_{lens}r} + \sqrt{\left(\frac{R_{lens}}{n_{lens}r}\right)^2 - 1} \qquad (1)$$

The refractive index of the lens changes from unity at its edge to infinity at its center. Our goal is to intersect a waveguide with the Eaton lens and consequently bend the light. To find the light ray trajectories, we place an array of point sources inside the waveguide's core and on its edges. The light ray trajectories in the Eaton lens with a radius of $R_{lens}$=10.5 μm are shown in Fig. 1(a). The light rays that correspond to the point sources located on the edges of the waveguide are bolded in this figure. The refractive index of the Eaton lens given by Eq. (1) is one at its edge which should match the refractive index of the waveguide's core to avoid reflection. Hence, the numerically calculated $n_{lens}$ should be multiplied by $n_{core}$. In Fig. 1(a), we limited the refractive index of the lens to 4.4 so it may be easier for the reader to distinguish the refractive index distribution of the lens. Since the light rays only pass thought certain part of the lens, truncation of the lens to this part does not degrade the performance of the lens as shown in Fig. 1(b), however, this truncation reduces the footprint of the bend. Moreover, the truncated lens improves the performance of the bend since it is surrounded by the cladding material and therefore acts as a waveguide. The bending radius is $R_{bend}$=18.4 μm for the designed bent waveguide. The performance of the bend without lens is shown in Fig. 2 (a). The propagating modes in the low-index contrast waveguide ignore the sharp bend and travel in a straight path leading to complete leakage of the mode's energy to outside of the bend. The complete Eaton lens bends the light by 90° but it is not successful in guiding the modes towards the exiting straight waveguide (Fig. 2(b)). However, as shown in Fig. 2(c), truncated Eaton lens effectively confines light inside the bent waveguide. As discussed previously, the cladding material surrounds the truncated Eaton lens which helps the lens to confine the light similar to a waveguide.

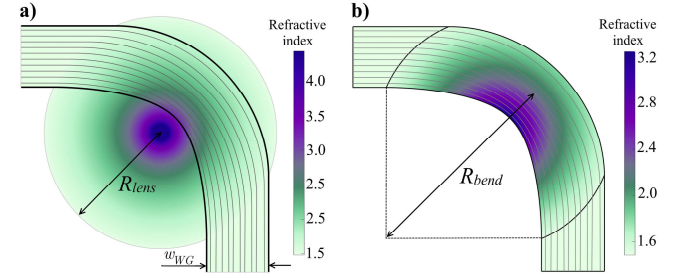

Fig. 1. a) Intersection of a waveguide with the Eaton lens. The rays, emitted from an array of point sources, are bended by the lens. The rays corresponding to the edges of the waveguides are bolded. b) The Eaton lens is truncated to improve the bend's performance and reduce its footprint.

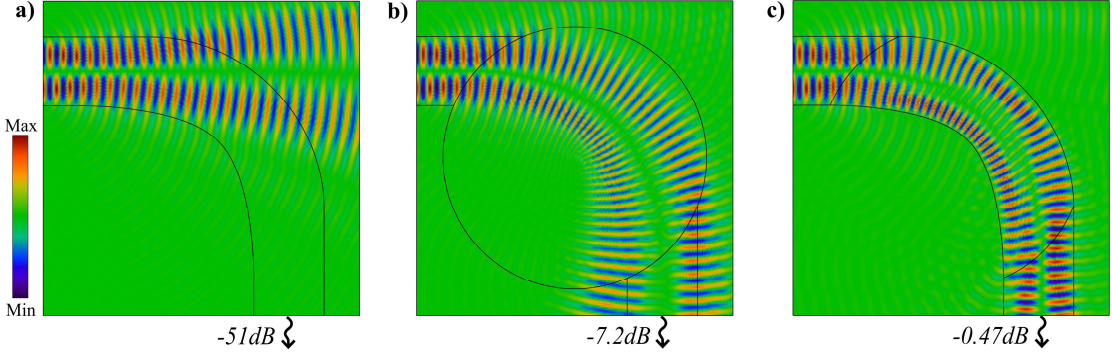

Fig. 2. The TM$_1$ mode's propagation through the polymeric waveguide bend a) without Eaton lens, b) with complete Eaton lens, and c) with truncated Eaton lens at 1550 nm. Truncation of the Eaton lens reduces the bending loss considerably.

## 3. Multilayered Eaton lens

Multilayered structures based on their interference effects have many applications such as antireflection coatings, Bragg mirrors, and filters [21-24]. In these structures, the thickness of the layers, $d$, is comparable to the wavelength of the light, $\lambda$. On the other hand, when the thickness of the layers is much smaller than the wavelength of the light, the interference effects are negligible. Therefore, in the long-wavelength limit, the multilayer structure can be regarded as a homogenous medium [25]. Multilayer structures as homogenous media have been utilized in various devices and applications such as hyperlenses, invisibility cloaks, couplers, and waveguide crossings [26-30]. Ensuring that $d \ll \lambda$, the effective medium theory (EMT) can be used to calculate the effective refractive index of the multilayer structure [31] provided that a sufficient number of layers exist. When the inclusion layers are arranged parallel to the electric field of the wave, the effective refractive index of the two-phase multilayer can be approximated by [32]:

$$n_{eff,TM}^2 = f_{inc}n_{inc}^2 + (1-f_{inc})n_{host}^2 \qquad (2)$$

where $f_{inc}$ is the fraction of the total volume occupied by inclusions. The $n_{host}$, $n_{inc}$, and $n_{eff,TM}$ are the refractive indices of the host, inclusion, and effective medium for transverse magnetic (TM) mode, respectively. The host material should have the same refractive index as the waveguide's core to minimize reflections ($n_{host}=n_{core}$). And inclusion layers are considered to be silicon ($n_{inc}=n_{Si}$=3.45). The simplest way to develop a computational model is to form a concentric cylindrical multilayer structure composed of alternating silicon (Si) and polymer layers for the whole Eaton lens. Afterwards, part of the rings intersecting the truncated Eaton lens are kept while the remaining parts of the rings are eliminated. We divide the circular lens into cylindrical layers with equal width, $\Lambda$, and the width of the inclusion layer in the $i$-th layer, $d_i$, is given by [30]:

$$d_i = \frac{n_{eff,TM}^2 - n_{host}^2}{n_{inc}^2 - n_{host}^2}\Lambda \qquad (3)$$

where $n_{eff,TM}$ is the average refractive index of that layer. The design concept is illustrated in Fig. 3(a) where the period of the structure ($\Lambda$) and width of inclusion layers are also specified. Only a quarter of the annular rings is displayed in this figure. The implemented ring-based multilayer structure for the truncated Eaton lens is shown in Fig. 3(b). The inclusion layers near the interface of the straight and bent waveguide are very thin and therefore their fabrication is very challenging. Inspired by graded photonic crystals, we limited the width of inclusion layers to 35 nm as shown in Fig. 3(c). In order to control the minimum width of the inclusion layers, each annular ring is divided into smaller annular sectors. Initially, the distance between the centers of two consecutive sectors is set to Λ. Then the arc length of the inclusion sectors with given width (35 nm) is calculated. When the inclusion sectors' arc length is smaller than 35nm, the distance between the sectors (Λ) are gradually increased to reach a desired arc length. With this method, the width of the inclusion layers in the annular sectors (Fig. 3(c)) increases compared to the annular rings of Fig. 3(b). In this paper, we focus on multilayer implementation of the GRIN medium but it can also be implemented by GPC structure [33, 34]. For the TE mode, the inclusion layers are arranged perpendicular to the electric field, the effective refractive index, $n_{eff,TE}$, is calculated by [30]:

$$n_{eff,TE}^2 = \frac{n_{inc}^2 n_{host}^2}{f_{inc} n_{host}^2 + (1-f_{inc}) n_{inc}^2} \quad (4)$$

Due to the form birefringence in multilayer structures, a multilayer bend can either support TM or TE mode.

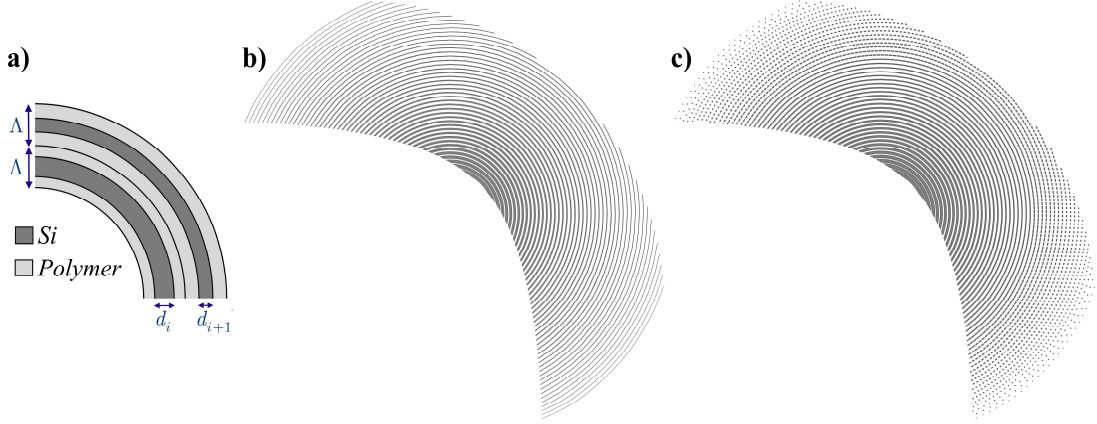

Fig. 3. a) Part of the annular rings with a period of Λ and inclusion layers with widths of $d_i$ and $d_{i+1}$ calculated by Eq. (3). The host and inclusion layers are considered to be polymer and Si, respectively. Implementation of the truncated Eaton lens by multilayer structure b) without limiting the width of inclusion layers and c) with limiting the inclusion layers' width to 35 nm. The period of structure is Λ=182 nm for both implementations. The host material is not shown in these implementations.

## 4. Numerical simulation and discussion

The two-dimensional (2D) finite-element method (FEM) was employed to evaluate the performance of the designed bent waveguide. We designed a waveguide bend with a radius of 18.4 μm for a 5.5 μm-wide polymeric waveguide. However, our method can be expanded to design a different bending radius for low-index waveguides with different width by changing the radius of the Eaton lens. For the given waveguide width, the minimum radius of the Eaton lens should be about two times of the waveguide width. Therefore, the minimum bending radius that can be designed by our method depends on the width of the waveguide. The method presented in this article is suitable for designing low-index waveguide bends. On the other hand, designing a high-index contrast waveguide bend by our method results in large bending radius. This is due to the fact that in our method, the refractive index of the lens is multiplied by $n_{core}$. Therefore, as $n_{core}$ increases the radius of the lens should be increased to avoid infeasible refractive indices for implementation of the bend. For instance, the bend designed for a waveguide with the effective refractive index of 2.2 results in a bending radius of approximately 55 μm. However, high-index contrast waveguides such as Si waveguides can be bent with small bending radii [11, 12, 15, 35, 36].

Based on the procedure described in the previous section, three implementations of the truncated Eaton lens with different periods are presented with periods of Λ=182, 194, and 207 nm. In all of these implementations, the minimum width of inclusion layers is limited to 35 nm. For Λ=182 nm, the electric field distribution of the $TM_0$ and $TM_1$ modes are shown in Fig. 4. The transmission values for these modes are also displayed in this figure for a wavelength of 1550 nm. As expected the bend loss is slightly higher in $TM_1$ mode compared to $TM_0$.

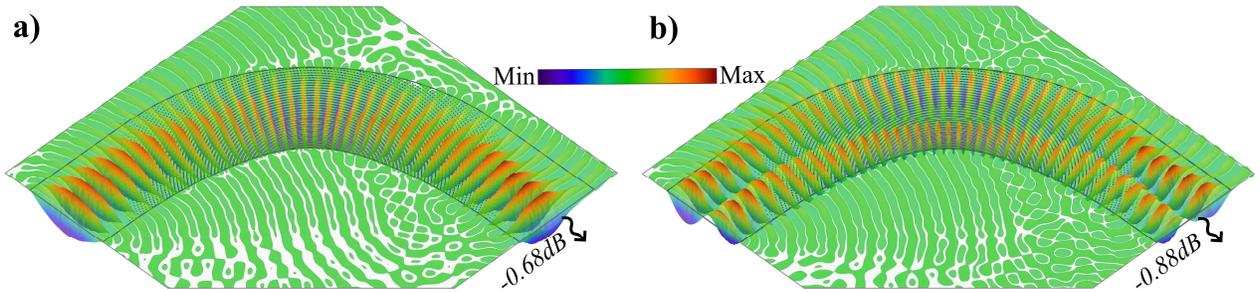

Fig. 4. Propagation of a) $TM_0$ and b) $TM_1$ modes at 1550nm through the waveguide bend implemented by multilayer Eaton lens (Λ=182 nm).

The bend loss of $TM_0$ mode for the ideal lens and multilayered lenses with different periods are displayed in Fig. 5. The average bend loss in the C-band is 0.69, 0.93, and 1.41 dB for the periods of Λ=182, 194, and 207 nm, respectively. Obviously, the bend loss increases as the period of structure increases. As the wavelength of the light increases, the structure seems more homogenous to the light and consequently the characteristics of the multilayer structure is closer to the ideal lens. For the structure with Λ=182 nm, the bend loss of TM$_0$ mode is lower than 1 dB in the whole range of 1470-1675 nm range.

In Fig. 6 the bend losses of the ideal lens and multilayer structures are shown for the TM$_1$ mode. The average bend loss in the C-band is 0.85, 1.28, and 2.36 dB for the periods of Λ=182, 194, and 207 nm, respectively. The bend loss is lower than 1 dB in 1520-1675 nm range for Λ=182 nm. The bend loss for the TM$_1$ mode is higher than the TM$_0$ mode due to the fact that for TM$_1$, the light is more sensitive to the structure's period. As the wavelength of light decreases, the bend loss of TM$_1$ mode increases with a higher rate compared to the TM$_0$ mode.

A possible method of implementing the proposed multilayer structure is to use electron-beam lithography combined with inductively coupled etching to fabricate Si layers on the substrate [37]. The Si layers can also be formed by nanoimprint lithography technique [38]. The nanoimprint lithography can also be used to fill the spaces between the Si layers with polymer [37]. To account for the inevitable fabrication imperfections in the performance of the designed bend, random deviations are introduced to the width of the Si layers. The numerical simulations show that a 10% random deviation, in the structure with Λ=182 nm, results in an excess bend loss of up to 0.5 dB in the C-band.

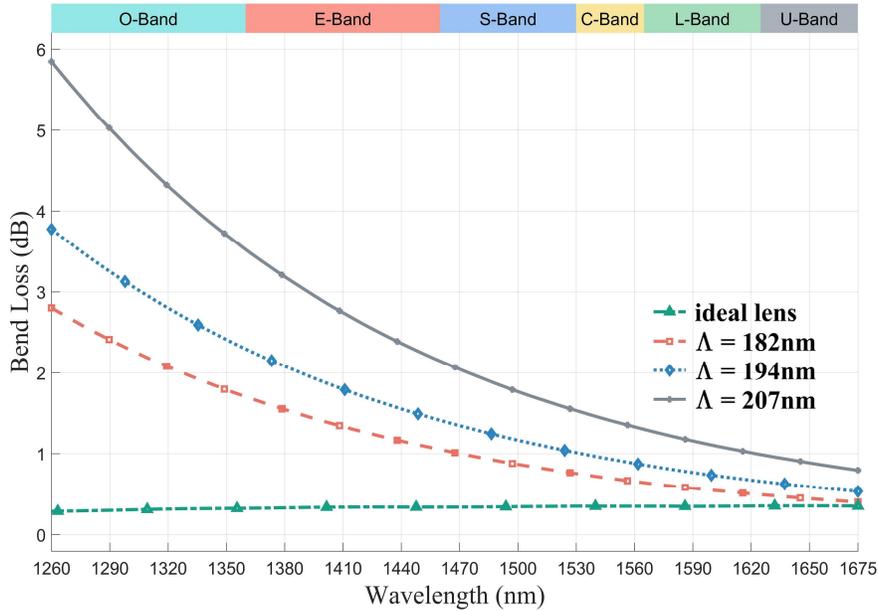

Fig. 5. The bend loss of TM$_0$ mode for the ideal and multilayered Eaton lenses with different periods.

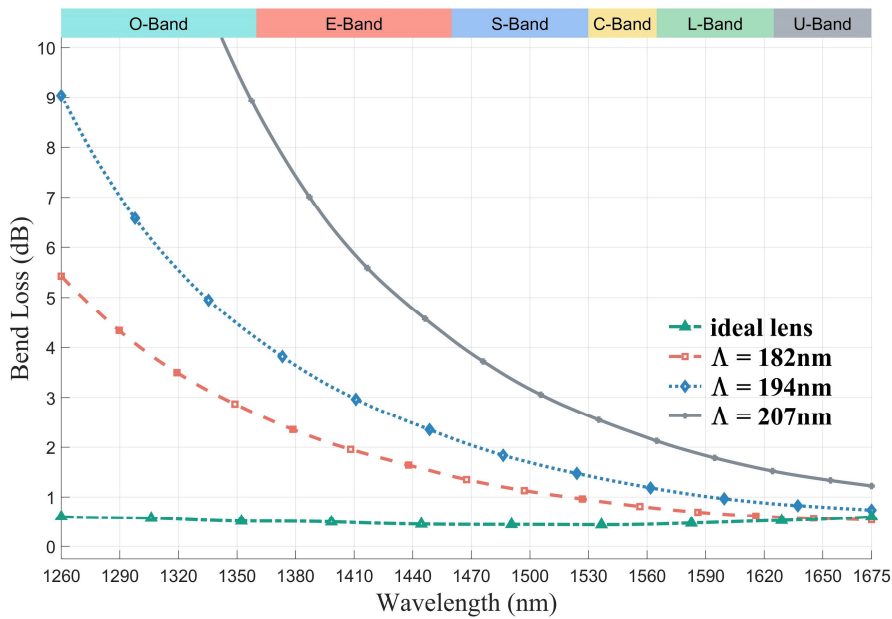

Fig. 6. The bend loss of TM$_1$ mode for the ideal and multilayered Eaton lenses with different periods.

A comparison is made between the bend designed in this article (Λ=182 nm case) and structures proposed at references [3-7] and the result is presented in Table 1. References [3, 5, 6] report 90° bend while reference [7] reports a 180° bend, and reference [4] reports a 45° bend. The reported polymeric multimode waveguide bends in references [3, 5, 6] have radii in the millimeter range. It should be noted that the width of the waveguides in these references is 50 μm. Scaling up our design to a 50 μm-wide waveguide results in a bending radius of approximately 180 μm which is still much smaller than the mentioned studies. It must be noted that the waveguides considered in this article have a much lower index contrast ($\Delta = n_{core}/n_{cladding} - 1$) compared to [3, 5, 6] which is much more difficult to bend with low loss. Moreover, as discussed in [6], by increasing the width of waveguide, the bending loss decreases.

Considering all these facts, it can be concluded that our design has a lower bending loss compared to [3, 5, 6]. Reference [4] reports the lowest bend loss but it has a large footprint of 63×63 μm$^2$. Moreover, it reports a 45° bend so for bending the waveguide by 90° it needs two 45° bends, therefore, it occupies even larger space for a 90° bend. We also compare a few silicon-on-insulator (SOI) waveguide bends in Table 1. Finally, it is worth noting that references discussed above, reported experimental results whereas we report numerical results. However, our results show that the presented method can be effectively employed to reduce the bending radius and footprint effectively and consequently, the footprint of PLCs incorporating many bends can be reduced considerably.

Table 1. Comparison of waveguide bends

| Ref. | Waveguide type | Bending mechanism | Waveguide width ($\mu m$) | Index contrast ($\Delta$) | Bending radius | Bend loss ($dB$) | Bandwidth ($nm$) | Multimode bend reported |
|---|---|---|---|---|---|---|---|---|
| [3] | Polymeric | - | 50 | 1.33% | 8 mm | 1 | - | Yes |
| [4] | Polymeric | air trench mirror | 3.6 | 0.13% | - | 0.16 | 1480-1580 | No |
| [5] | Polymeric | tapered waveguide | 50 | 1.94% | 14 mm | 0.78 | - | Yes |
| [6] | Polymeric | - | 50 | 1.94% | 13.5 mm | 0.74 | - | Yes |
| [7] | Polymeric | air trench & offset | 6 | 0.69% | 1.5 mm | 3 | - | No |
| this work | Polymeric | Eaton lens | 5.5 | 0.13% | 18.4 μm | 1 | 1520-1675 | Yes |
| [11] | SOI | Mode converters | 2.5 | 94% | 30 μm | 1 | 1520-1600 | Yes |
| [15] | SOI | Inverse design | 2 | 180% | 1 μm | 1 | 1530-1570 | Yes |
| [35] | SOI | Subwavelength grating | 1.2 | 94% | 10 μm | 0.5 | 1500-1600 | Yes |

## 5. Conclusion

Reducing the bending radius of multimode waveguide bends while preserving the loss to low levels is crucial to reduce the footprint of optical components and integrated circuits. In this article, a polymeric multimode waveguide bend is designed based on the truncation of the Eaton lens which is implemented as a ring-based multilayer structure and its characteristics is numerically evaluated by the finite element method. The designed waveguide bend had a bending radius of 18.4 μm for a waveguide width of 5.5 μm and a low index contrast of only 0.13%. The average bend losses of 0.69 and 0.87 dB are achieved for the $TM_0$ and $TM_1$ modes in the C-band, respectively. The results indicate that compared to the previous designs, the polymeric waveguide bend presented in this article has a much smaller footprint without any significant increase in the bending loss value.